\documentstyle[multicol,eqsecnum,aps,psfig]{revtex}
\renewcommand{\narrowtext}{\begin{multicols}{2}
\global\columnwidth20.5pc\noindent}
\renewcommand{\widetext}{\end{multicols}
\global\columnwidth42.5pc}
\begin{document}
\draft
\preprint{14 August 2003}
\title{Recent Progress of the Low-Dimensional Spin-Wave Theory}
\author{Shoji Yamamoto}
\address{Division of Physics, Hokkaido University,
         Sapporo 060-0810, Japan}
\date{Received 14 August 2003}
\maketitle
\begin{abstract}
A modified spin-wave theory is developed and applied to low-dimensional
quantum magnets.
Double-peaked specific heat for one-dimensional ferrimagnets, nuclear
spin-lattice relaxation in ferrimagnetic chains and clusters, and thermal
behavior of Haldane-gap antiferromagnets are described within the scheme.
Mentioning other bosonic and fermionic representations as well, we
demonstrate that {\it spin waves are still effective in low dimensions}.
\end{abstract}
\narrowtext

\section{Introduction}

   It may be spin waves that first come into our mind when we discuss
low-energy magnetic excitations.
The spin-wave concept$-$local excitations can be stabilized by recovering
the original symmetry$-$is widely valid in physics.
The ferromagnetic spin-wave theory was initiated by Bloch \cite{B206} and
quantum-mechanically formulated  by Holstein and Primakkof \cite{H1098}.
Dyson \cite{D1217} considered correlations between spin waves
\cite{O117,K25,K1384}.
Anderson \cite{A694} and Kubo \cite{K568} developed the scheme to
antiferromagnets.
Such theoretical arguments motivated numerous experiments
\cite{G122,W1357,N1929,D433}.
Nowadays the spin-wave picture still stimulates many researchers to
further explorations.

   On the other hand, we may have a vague but fixed idea that spin waves
are no more valid in low dimensions where quantum as well as thermal
fluctuations break down any magnetic long-range order.
Low-dimensional materials were synthesized one after another, but the
spin-wave theory was not able to play any effective role in studying their
quantum behavior.
While dynamic properties can be potentially revealed in terms of spin
waves \cite{M23,P398}, their application \cite{N354,K357} was restricted
to sufficiently low temperatures assuming the onset of the
three-dimensional long-range order.
In such circumstances, (C$_6$H$_{11}$NH$_3$)CuCl$_3$ \cite{W1055}, which
turned out a model compound for one-dimensional ferromagnets, stimulated
public interest in low-dimensional thermodynamics.
Numerically solving the thermodynamic Bethe-ansatz equations for the
one-dimensional Heisenberg model
\begin{equation}
   {\cal H}
   =J\sum_{l=1}^L
     \mbox{\boldmath$S$}_l\cdot\mbox{\boldmath$S$}_{l+1}\,,
\end{equation}
Takahashi and Yamada \cite{T2808} found that the spin-$\frac{1}{2}$
ferromagnetic specific heat and magnetic susceptibility are expanded as
\begin{eqnarray}
   &&
   \frac{C}{Lk_{\rm B}}
      = 0.7815t^{\frac{1}{2}}-2.00t+3.5t^{\frac{3}{2}}
      +O(t^2)\,,
   \label{E:MSWLTSE-C}
   \\
   &&
   \frac{\chi J}{L(g\mu_{\rm B})^2}
      = 0.04167t^{-2}-0.145t^{-\frac{3}{2}}+0.17t^{-1}
      +O(t^{-\frac{1}{2}})\,,
   \nonumber \\
   &&
   \label{E:MSWLTSE-S}
\end{eqnarray}
at low temperatures, where $t=-k_{\rm B}T/J$ ($J<0$).
Schlottmann \cite{S2131} also obtained similar results.
These findings not only settled the long-argued problem of the
low-temperature behavior of one-dimensional ferromagnets
\cite{B640,B1272,K807,C297,L3108} but also motivated revisiting the
subject by the spin-wave scheme.

   The conventional spin-wave theory \cite{H1098} applied to
one-dimensional ferromagnets gives no quantitative information on the
susceptibility but reveals the correct leading term of the specific heat
at low temperatures.
Spin waves still look valid in low dimensions.
The difficulty of diverging magnetization originates in uncontrollable
thermal excitations within the conventional spin-wave scheme.
Then we may have an idea of constraining spin waves to keep zero
magnetization \cite{T233,T168}, which is reasonable for isotropic magnets.
The thus-calculated thermal properties of spin-$S$ Heisenberg chains,
\begin{eqnarray}
   &&
   \frac{C}{Lk_{\rm B}}
      = \frac{3}{8S^{1/2}}\frac{\zeta(\frac{3}{2})}{\sqrt{\pi}}t^{1/2}
      - \frac{1}{2S^2}t
      + \frac{15}{32S^{7/2}}
   \nonumber \\
   &&\qquad\times
        \left[
         \frac{S^2}{4}\frac{\zeta(\frac{5}{2})}{\sqrt{\pi}}
        -\frac{\zeta(\frac{1}{2})}{\sqrt{\pi}}
        \right]t^{3/2}
      + O(t^2)\,,
   \label{E:MSWferro-C}\\
   &&
   \frac{\chi J}{L(g\mu_{\rm B})^2}
      = \frac{2S^4}{3} t^{-2}
      - S^{5/2}\frac{\zeta(\frac{1}{2})}{\sqrt{\pi}}t^{-3/2}
   \nonumber \\
   &&\qquad\qquad
      + \frac{S}{2}
        \left[
         \frac{\zeta(\frac{1}{2})}{\sqrt{\pi}}
        \right]^2 t^{-1}
      + O(t^{-1/2})\,,
   \label{E:MSWferro-S}
\end{eqnarray}
coincide very well with Eqs. (\ref{E:MSWLTSE-C}) and (\ref{E:MSWLTSE-S})
in the case of $S=\frac{1}{2}$.
This scheme was further applied to two-dimensional square-lattice
\cite{T2494,H4769} and frustrated \cite{C7832,D13821} antiferromagnets
and is now generically referred to as the {\it modified spin-wave theory}.

   The spin-wave theory thus expanded toward lower dimensions and higher
temperatures.
However, many systems are still left for spin waves to explore.
There are few attempts \cite{R2589} at describing one-dimensional
antiferromagnets in terms of spin waves beyond the naivest application
\cite{A694,K568}.
In the case of antiferromagnets, one and two dimensions are essentially
distinguished.
The N\'eel order exists in Heisenberg square-lattice antiferromagnets
\cite{M1}, whereas no long-range order in Hisenberg liner-chain
antiferromagnets.
Correspondingly, spin waves claim that the ground-state sublattice
magnetizations are convergent in two dimension but divergent in one
dimension.
Hence spin waves have been considered to be useless for one-dimensional
quantum antiferromagnets.
It is also unfortunate that poor progress has been made in spin-wave
treatments of one-dimensional quantum ferrimagnets, where interesting
{\it mixed magnetism} \cite{Y1024} lies.

   Thus motivated, we first construct a modified spin-wave theory for
one-dimensional ferrimagnets with various lattice structures.
The thermal and dynamic properties are fully revealed.
Then we proceed to ferrimagnetic clusters, which is, to my knowledge, the
first application of spin waves to zero dimension.
Finally we demonstrate a modified spin-wave description of spin-gapped
antiferromagnets and compare it with a Schwinger-boson mean-field
approach.

\section{One-Dimensional Quantum Ferrimagnets}

   Let us start our discussion from one-dimensional Heisenberg
ferrimagnets composed of alternating spins $S$ and $s$:
\begin{equation}
   {\cal H}
      =J\sum_{n=1}^N
        \left(
         \mbox{\boldmath$S$}_{n} \cdot \mbox{\boldmath$s$}_{n}
        +\mbox{\boldmath$s$}_{n} \cdot \mbox{\boldmath$S$}_{n+1}
        \right)\,.
   \label{E:HSs}
\end{equation}
When we define quantum spin reduction $\tau$ as $M_S=N(S-\tau)$ and
$M_s=-N(s-\tau)$, where $M_S$ and $M_s$ are the ground-state sublattice
magnetizations, spin waves show that \cite{B3921}
\begin{equation}
   \tau=\int_0^\pi
    \Biggl[
     \frac{S+s}{\sqrt{(S-s)^2+4Ss\sin^2\frac{k}{2}}}-1
    \Biggr]\frac{{\rm d}k}{2\pi}\,.
   \label{E:tau}
\end{equation}
This integral is divergent at $k=0$ in the case of $S=s$, which is the
problem of {\it infrared-diverging magnetization}.
Ferrimagnets, however, get rid of this difficulty.
Equation (\ref{E:tau}) gives $\tau\simeq 0.305$ at
$(S,s)=(1,\frac{1}{2})$, for example, while the correct value turns out
$0.207(1)$ \cite{K3336}.
The spin-wave scheme is trivially valid for large spins.
Considering that the quantity $S-s$ suppresses the divergence in
Eq. (\ref{E:tau}), the spin-wave scheme is better justified with increasing
$S/s$ as well as $Ss$ \cite{Y211}. 

   For ferrimagnets, because of no quantum divergence, we can consider
modifying the conventional spin-wave scheme either according to the
Takahashi scheme \cite{T168}, that is, in diagonalizing the Hamitonian, or
in the stage of constructing thermodynamics.
Since the original Takahashi scheme ends in a poorer description of
thermal properties, here we take the latter approach.
Introducing bosonic operators for the spin deviation in each sublattice
via the Holstein-Primakoff transformation
\begin{equation}
   \left.
   \begin{array}{lll}
    S_n^+=\sqrt{2S-a_n^\dagger a_n}\ a_n\,,&
    S_n^z=S-a_n^\dagger a_n\,,\\
    s_n^+=b_n^\dagger\sqrt{2s-b_n^\dagger b_n}\ ,&
    s_n^z=-s+b_n^\dagger b_n\,,
   \end{array}
   \right.
   \label{E:HP}
\end{equation}
we expand the Hamiltonian (\ref{E:HSs}) with respect to $1/S$
($1/4S$ more strictly \cite{W433}) as
\begin{equation}
   {\cal H}=-2NJSs+{\cal H}_1+{\cal H}_0+O(S^{-1})\,,
   \label{E:Hexp}
\end{equation}
where we have treated $S$ and $s$ as $O(S)=O(s)$.
We consider the two-body term ${\cal H}_0$ of $O(S^0)$ to be a
perturbation to the one-body term ${\cal H}_1$ of $O(S^1)$.
Via the Fourier transformation
\begin{equation}
   \left.
   \begin{array}{ccc}
    a_k^\dagger
    &=&{\displaystyle\frac{1}{\sqrt{N}}}
       {\displaystyle\sum_n}{\rm e}^{-2{\rm i}ak(n-1/4)}a_n^\dagger\,,\\
    b_k^\dagger
    &=&{\displaystyle\frac{1}{\sqrt{N}}}
       {\displaystyle\sum_n}{\rm e}^{-2{\rm i}ak(n+1/4)}b_n^\dagger\,,\\
   \end{array}
   \right.
   \label{E:Fourier}
\end{equation}
with $a$ being the lattice constant and then the Bogoliubov transformation
\begin{equation}
   \left.
   \begin{array}{ccc}
    \alpha_k^\dagger
    &=&a_k^\dagger\mbox{cosh}\theta_k\
     + b_k        \mbox{sinh}\theta_k\,,\\
    \beta_k^\dagger
    &=&a_k        \mbox{sinh}\theta_k
     + b_k^\dagger\mbox{cosh}\theta_k\,,
   \end{array}
   \right.
   \label{E:Bogliubov}
\end{equation}
with $\mbox{tanh}2\theta_k=2\sqrt{Ss}\,\mbox{cos}(ak)/(S+s)$,
we obtain a compact expression
\begin{equation}
   {\cal H}_i
   =E_i+\sum_k
        \left[
         \omega_i^-(k)\alpha_k^\dagger\alpha_k
        +\omega_i^+(k)\beta_k^\dagger\beta_k
        \right]\,,
   \label{E:HSW}
\end{equation}
\widetext
\begin{figure}
\centerline
{\mbox{\psfig{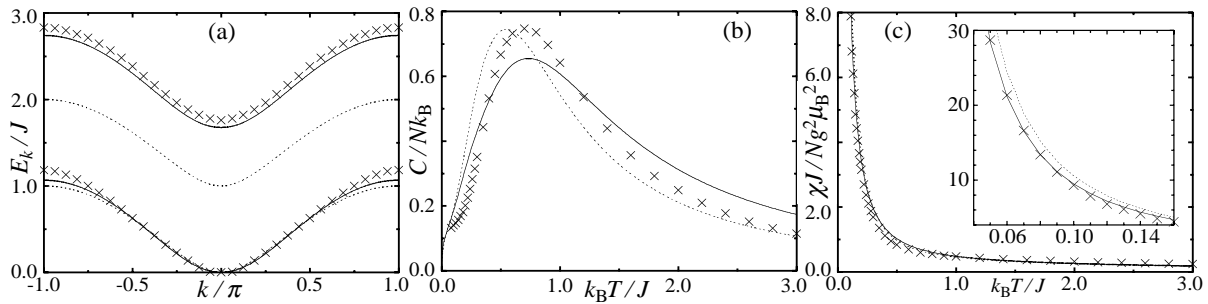}}}
\vspace*{2mm}
\caption{Dispersion relations of the ferromagnetic and antiferromagnetic
         elementary excitations (a) and temperature dependences of the
         specific heat (b) and the magnetic susceptibility (c) for the
         $(S,s)=(1,\frac{1}{2})$ ferrimagnetic Heisenberg chain.
         The linear (dotted lines) and interacting (solid lines) spin-wave
         calculations are compared with quantum Monte Carlo
         estimates.}
\label{F:SsFT}
\end{figure}
\narrowtext
\noindent
where $E_i$ and $\omega_i^\pm$ are the $O(S^i)$ quantum corrections to the
dispersion relations.
There exist two branches of spin waves in ferrimagnets, one of which
reduces the ground-state magnetization and is thus of ferromagnetic
aspect, while the other of which enhances the ground-state magnetization
and is thus of antiferromagnetic aspect.
The ferromagnetic and antiferromagnetic spin waves convincingly exhibit
a quadratic dispersion and a gapped spectrum, respectively, as is shown
in Fig. \ref{F:SsFT}(a).

   While the formulation is detailed in Ref. \cite{Y14008}, we here
mention the core idea of our ferrimagnetic modified spin-wave theory$-$how
to control the number of bosons.
The zero-magnetization constraint, which works well in ferromagnets, is
not relevant to antiferromagnetically coupled spins, keeping constant
the difference between the numbers of the sublattice bosons instead of the
sum of them.
Then we may consider that the staggered magnetization should be kept
constant and introduce an alternative constraint as
\begin{equation}
   \sum_{\sigma=\pm}\sum_k
   \frac{(S+s)\bar{n}_k^\sigma}{\sqrt{(S-s)^2+4Ss\sin^2(ak)}}
   =(S+s)N\,,
   \label{E:Mst:}
\end{equation}
where $\bar{n}_k^-=\langle\alpha_k^\dagger\alpha_k\rangle$ and
      $\bar{n}_k^+=\langle\beta_k^\dagger \beta_k\rangle $.
Equation (\ref{E:Mst:}) claims that {\it the thermal fluctuation should
cancel the staggered magnetization}.
The thus-calculated thermal quantities are shown in Fig. \ref{F:SsFT}.
The ferromagnetic features of ferrimagnets at low temperatures, the
$T^{1/2}$-vanishing specific heat and the $T^{-2}$-diverging
susceptibility, are reproduced well.
The inset in Fig. \ref{F:SsFT}(c) shows that the interacting modified
spin-wave description is highly precise at sufficiently low temperatures.
As for the specific heat, the inclusion of the spin-wave interactions
corrects the position of the Schottky peak, which is consistent with
the observations of Fig. \ref{F:SsFT}(a).

   There are many model compounds for the Hamiltonian (\ref{E:HSs}).
Kahn {\it et al.} \cite{K782} systematically synthesized
spin-$(S,\frac{1}{2})$ quasi-one-dimensional ferrimagnets composed of two
kinds of transition metals.
More complicated alignments of mixed spins \cite{O5221,O8067} were also
synthesized and investigated in terms of modified spin waves.

\section{Double-Peaked Specific Heat}

\begin{figure}
\centerline
{\mbox{\psfig{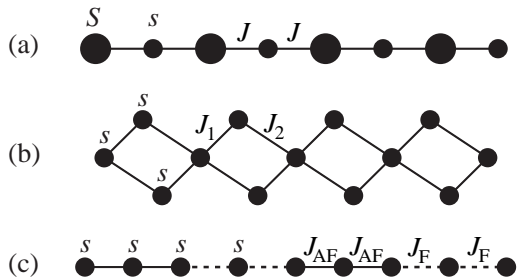}}}
\vspace*{3mm}
\caption{Schematic representations of bimetallic chain compounds (a),
         trimeric intertwining double-chain compounds (b), and tetrameric
         bond-alternating chain compounds (c), where smaller and larger
         bullet symbols denote spins $s=\frac{1}{2}$ and $S>\frac{1}{2}$,
         while solid and dashed segments mean antiferromagnetic and
         ferromagnetic exchange couplings between them, respectively.}
\label{F:ferri}
\end{figure}

   The ferromagnetic and antiferromagnetic excitations coexistent in
one-dimensional ferrimagnets lead to the mixed features of the specific
heat: the $T^{1/2}$ initial behavior but the Schottky-type peak at mid
temperatures.
More interesting energy structure may be expected in ferrimagnets.
Weak magnetic field applied to ferrimagnets indeed causes a double-peaked
specific heat \cite{M5908}.
When the antiferromagnetic gap lies far apart from the lower-lying
ferromagnetic band, there is a possibility of such observations in
isotropic systems without any field and anisotropy.
In the case of alternating-spin chains illustrated in
Fig. \ref{F:ferri}(a), the linear spin-wave theory suggests the criterion
as $S\gg 2s$.
A series of bimetallic chain compounds \cite{K782} is interesting in this
context, but even a combination of Mn ($S=\frac{5}{2}$) and Cu
($s=\frac{1}{2}$) turns out not enough for a double-peaked structure of
the specific heat \cite{D10992}.

\begin{figure}
\centerline
{\mbox{\psfig{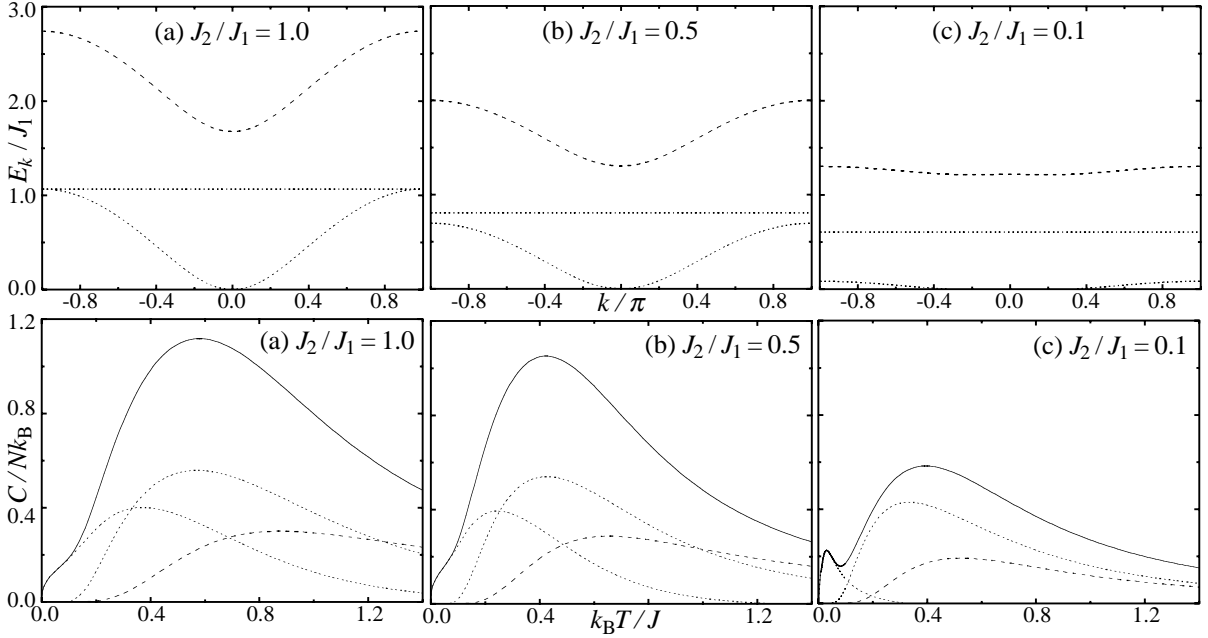}}}
\vspace*{2mm}
\caption{The interacting spin-wave calculations of the dispersion
         relations (the upper three) and the specific heat (the lower
         three) of the trimeric chains illustrated in Fig.
         \ref{F:ferri}(b), where individual contributions of the
         ferromagnetic and antiferromagnetic spin waves are
         distinguishably shown by dotted and dashed lines, respectively,
         the sum of which is equal to the total drawn by solid lines.}
\label{F:J1J2C}
\end{figure}

   Metals of different kinds are not necessary to ferrimagnets.
Caneschi {\it et al.} \cite{C56} synthesized hybrid ferrimagnets
comprising metals and organic radicals.
There are further solutions to ferrimagnets.
Figures \ref{F:ferri}(b) and \ref{F:ferri}(c) illustrate ferrimagnets of
topological origin \cite{D83,E4466}, which possibly exhibit a double-peaked
specific heat \cite{N214418}.
Figure \ref{F:J1J2C} shows the modified spin-wave calculations for the
spin-$\frac{1}{2}$ trimeric chains depicted in Fig. \ref{F:ferri}(b) with
particular emphasis on clarifying how an extra peak grows.
The present scheme has the advantage of visualizing each excitation mode
making a distinct contribution to the thermal behavior.
As the ratio $J_2/J_1$ moves away from unity, the system begins to divide
into small clusters and every excitation mode becomes less dispersive.
With a sufficiently small but finite ratio, an extra peak appears at low
temperatures.
The model compound Sr$_3$Cu$_3$(PO$_4$)$_4$ indeed exhibits a
double-peaked specific heat satisfying the condition $J_2/J_1\simeq 0.1$
\cite{D83}.
A double-peaked specific heat is possible for the tetrameric chains
depicted in Fig. \ref{F:ferri}(c) as well \cite{N214418}.
Extensive measurements on the model compound
Cu(3-Clpy)$_2$(N$_3$)$_2$ (3-Clpy $=$ $3$-chloropyridine) \cite{E4466,H30}
and its analogs are encouraged.

   Considering the difficulty of calculating thermodynamic properties at
very low temperatures by numerical tools such as quantum Monte Carlo and
density-matrix renormalization-group methods, the modified spin-wave
scheme can play an effective role in investigating low-dimensional magnets
as well.
We can clearly understand what type of fluctuations is relevant to the
phenomena of our interest in terms of spin waves.
Indeed any spin-wave description is trivially less quantitative in systems
with strong bond alternation and/or frustration \cite{Y13610}, but there
may be an idea that quantum phase transitions can be detected through the
breakdown of spin-wave ground states \cite{C7832,D13821}.
\widetext
\vspace*{2mm}
\narrowtext

\section{Nuclear Spin-Lattice Relaxation in Ferrimagnets}

   Next we consider observing ferrimagnets through nuclear spins.
In the case of a bimetallic chain compound
NiCu(pba)(H$_2$O)$_3$$\cdot$$2$H$_2$O
(pba $=$ $1,3$-propylenebis(oxamato)) synthesized by Kahn {\it et al.}
\cite{K782}, proton nuclei lying apart from electronic spins effectively
work as probes to illuminate the correlation between spins of different
kinds peculiar to ferrimagnets.
Considering the electronic-nuclear energy-conservation requirement, the
Raman process should play a leading role in the nuclear spin-lattice
relaxation.
The Raman relaxation rate is generally given by
\begin{eqnarray}
   &&
   \frac{1}{T_1}
    =\frac{4\pi\hbar(g\mu_{\rm B}\gamma_{\rm N})^2}
          {\sum_i{\rm e}^{-E_i/k_{\rm B}T}}
     \sum_{i,f}{\rm e}^{-E_i/k_{\rm B}T}
   \nonumber\\
   &&\ \times
     \big|
      \langle f|\sum_n(A_nS_n^z+B_ns_n^z)|i\rangle
     \big|^2
     \,\delta(E_f-E_i-\hbar\omega_{\rm N})\,,
\label{E:T1def}
\end{eqnarray}
where
$A_n$ and $B_n$ are the dipolar hyperfine coupling constants between
the nuclear and electronic spins in the $n$th unit cell,
$\omega_{\rm N}\equiv\gamma_{\rm N}H$ is the Larmor frequency of the
nuclei with $\gamma_{\rm N}$ being the gyromagnetic ratio, and the
summation $\sum_i$ is taken over all the electronic eigenstates
$|i\rangle$ with energy $E_i$.
Taking account of the significant difference between the electronic
and nuclear energy scales ($\hbar\omega_{\rm N}\alt 10^{-5}J$),
Eq. (\ref{E:T1def}) is expressed in terms of modified spin waves as
\cite{Y842}
\begin{eqnarray}
   &&
   \frac{1}{T_1}
   \simeq\frac{4\hbar(g\mu_{\rm B}\gamma_{\rm N})^2}{\pi J}
   \int_0^\pi{\rm d}k
   \frac{S-s}{\sqrt{(Ssk)^2+2(S-s)Ss\hbar\omega_{\rm N}/J}}
   \nonumber \\
   &&\qquad\times
   \bigl[
    (A{\rm cosh}^2\theta_k-B{\rm sinh}^2\theta_k)^2
     \bar{n}_k^-(\bar{n}_k^- +1)
   \nonumber \\
   &&\qquad
   +(A{\rm sinh}^2\theta_k-B{\rm cosh}^2\theta_k)^2
     \bar{n}_k^+(\bar{n}_k^+ +1)
   \bigr]\,,
   \label{E:T1final}
\end{eqnarray}
where $A$ and $B$ are the Fourier transforms of the coupling constants on
the assumption that their momentum dependence can be neglected \cite{F433}.
\vspace*{1mm}
\begin{figure}
\centerline
{\mbox{\psfig{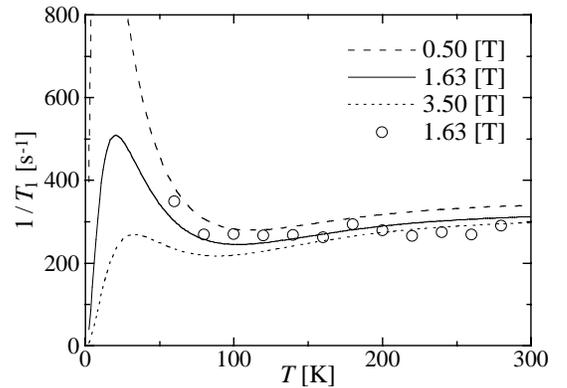}}}
\vspace*{2mm}
\caption{Temperature dependences of the proton spin-lattice relaxation
         rate with varying field for the $(S,s)=(1,\frac{1}{2})$
         ferrimagnetic model compound
         NiCu(pba)(H$_2$O)$_3$$\cdot$$2$H$_2$O.
         The interacting modified spin-wave calculations, where
         $A=1.8\times 10^{-2}$ $\mbox{\AA}^{-3}$ and
         $B=5.9\times 10^{-2}$ $\mbox{\AA}^{-3}$, are compared with
         measurements ($\circ$) [51].}
\label{F:SsT1Tex}
\end{figure}
\widetext
\begin{figure}
\centerline
{\mbox{\psfig{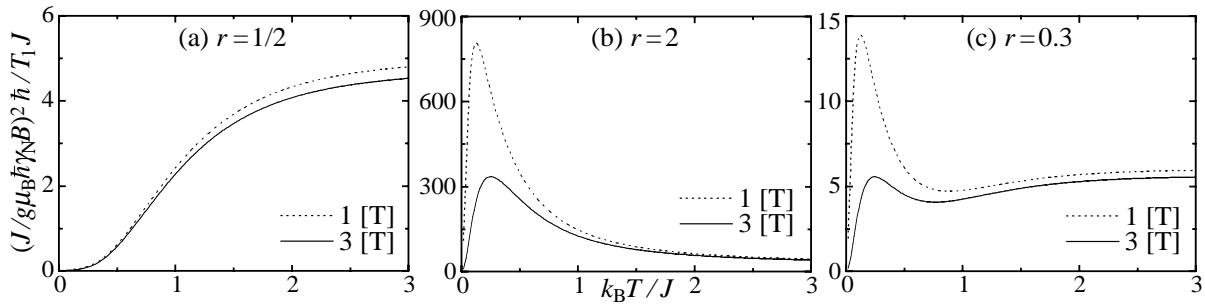}}}
\vspace*{2mm}
\caption{Temperature dependences of the nuclear spin-lattice relaxation
         rate with varying field for the $(S,s)=(1,\frac{1}{2})$
         ferrimagnetic Heisenberg chain calculated in terms of the
         interacting modified spin waves.}
\label{F:SsT1Tth}
\end{figure}
\vspace*{2mm}
\narrowtext

   In Fig. \ref{F:SsT1Tex} the thus-calculated relaxation rate
\cite{Y2324} is compared with experimental findings \cite{F433}.
We have determined the coupling constants so as to well reproduce the
observations under the condition that the probe protons are located closer
to Cu than Ni \cite{K782}.
Through the relations $A\sim d_S^{-3}$ and $B\sim d_s^{-3}$, where $d_S$
($d_s$) is the average distance between the protons and the Ni (Cu) site
in each unit, we obtain rough estimates
$d_S\simeq 3.8\,[\mbox{\AA}]$ and $d_s\simeq 2.6\,[\mbox{\AA}]$, which
are consistent with structural analyses \cite{K782}.
However, the calculated field dependence is much weaker than the
observations \cite{F433}.
The field dependence can be better interpreted by the direct process
\cite{Y1} than by the Raman process.
Since there are few nuclear-magnetic-resonance measurements on
one-dimensional ferrimagnets, we hope that further explorations will be
made from both chemical and physical points of view.

   Apart from particular materials, we discuss the ferrimagnetic
relaxation more generally.
When we compare Eq. (\ref{E:T1final}) with the expression of the
susceptibility-temperature product \cite{Y14008},
\begin{equation}
   \chi T
    =\frac{(g\mu_{\rm B})^2}{3k_{\rm B}}
     \sum_k\sum_{\sigma=\pm}
     \bar{n}^\sigma_k(\bar{n}^\sigma_k+1)\,,
   \label{E:chiTMSW}
\end{equation}
we find that the ferromagnetic and antiferromagnetic excitations
interestingly make varying contribution to $1/T_1$ according to the
location of probe nuclei.
Decreasing and increasing $\chi T$ with increasing temperature are
characteristic of ferro- and antiferromagnets, respectively, and
ferrimagnetic $\chi T$ exhibit a minimum.
As long as we observe ferrimagnetism through the susceptibility, it looks
like a uniform mixture of ferro- and antiferromagnetism.
On the other hand, nuclear spins can selectively extract ferro- and
antiferromagnetic features from ferrimagnets according to their location.
We show in Fig. \ref{F:SsT1Tth} the relaxation rate as a function of
$A/B\equiv r$.
The nuclei lie closer to larger (smaller) spins $S$ ($s$) for $r>1$
($r<1$).
Equation (\ref{E:T1final}) claims that ferromagnetic (antiferromagnetic)
spin waves can not mediate the nuclear spin relaxation at all for
$r=\mbox{tanh}^2\theta_k$ ($r=\mbox{coth}^2\theta_k$).
Considering the predominant contribution of $k\simeq 0$ to the
integration (\ref{E:T1final}) at low temperatures and weak fields
\cite{Y2324}, the nuclear spins correlate only to antiferromagnetic
(ferromagnetic) spin waves at $r\simeq s/S$ ($r\simeq S/s$).
Figures \ref{F:SsT1Tth}(a), \ref{F:SsT1Tth}(b), and \ref{F:SsT1Tth}(c)
indeed present antiferromagnetic, ferromagnetic, and mixed-magnetic
features, respectively, where vanishing rates at low temperatures in Figs.
\ref{F:SsT1Tth} (b) and \ref{F:SsT1Tth}(c) are due to the Zeeman
interaction and should therefore be distinguished from the intrinsic
properties.
Since antiferromagnetic spin waves are hardly excited at low temperatures,
{\it nuclear spins located so as to satisfy the condition
$r\sim(d_s/d_S)^3\simeq s/S$ exhibit extremely slow dynamics}.
\begin{figure}
\centerline
{\mbox{\psfig{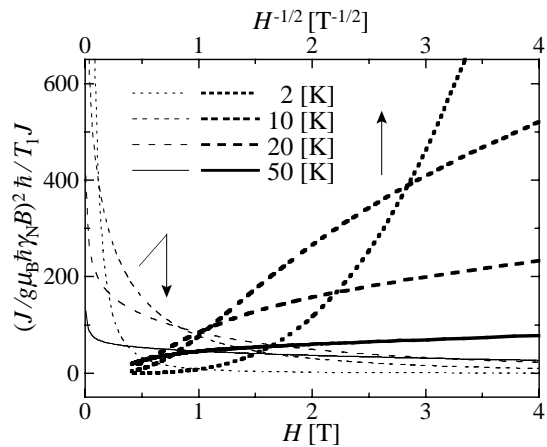}}}
\vspace*{2mm}
\caption{Field dependences of the nuclear spin-lattice relaxation rate
         with varying temperature for the $(S,s)=(1,\frac{1}{2})$
         ferrimagnetic Heisenberg chain calculated in terms of the
         interacting modified spin waves,
         where $J/k_{\rm B}=121$ $\mbox{K}$ and $r=0$.}
\label{F:SsT1Hth}
\end{figure}
\vspace*{2mm}

   Another interest in Eq. (\ref{E:T1final}) is its field dependence.
While it is usual for the relaxation rate to depend on an applied field
through the electronic Zeeman energy, here is further field dependence in
the prefactor.
This novel field dependence originates in the quadratic dispersion
relation of ferrimagnets and may arise from any nonlinear dispersion at
the band bottom in more general.
In Fig. \ref{F:SsT1Hth} we show the relaxation rate as a function of an
applied field, setting the parameters for 
NiCu(pba)(H$_2$O)$_3$$\cdot$$2$H$_2$O \cite{K782}. 
At low temperatures and moderate fields, $1/T_1$ behaves like
$\propto 1/\sqrt{H}$, where $\bar{n}_k^-$ exhibits a sharp peak at $k=0$
and thus the integration (\ref{E:T1final}) can be approximately replaced
by the $k=0$ contribution.
With increasing temperature, the predominant $k=0$ contribution smears and
the $k$ integration ends up with a {\it logarithmic field dependence}.
With increasing field, these unique observations are all masked behind the
overwhelming Zeeman effect.
The $1/\sqrt{H}$ dependence of our interest should be distinguished from
the diffusion-dominated dynamics \cite{H965,T2173,F11945}, which appears at
high temperatures originating from transverse spin fluctuations.
We hope that ferrimagnetic dynamics in a field will be more and more
studied from the experimental point of view.

\section{Nanoscale Molecular Magnets}

\begin{figure}
\centerline
{\mbox{\psfig{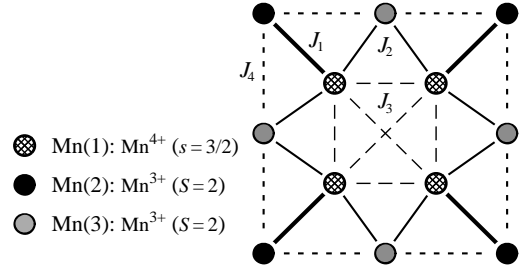}}}
\vspace*{2mm}
\caption{Schematic plot of the Mn$_{12}$ cluster.}
\label{F:Mn12}
\end{figure}

   Mesoscopic magnetism \cite{C661} is one of the hot topics in
materials science, where we can observe a dimensional crossover on the way
from molecular to bulk magnets \cite{H0544XX}.
Metal-ion magnetic clusters are thus interesting and among others is
[Mn$_{12}$O$_{12}$(CH$_3$COO)$_{16}$(H$_2$O)$_4$] \cite{L2042}
(hereafter abbreviated as Mn$_{12}$), for which quantum tunneling of the
magnetization \cite{F3830,T145} was observed for the first time.
Figure \ref{F:Mn12} illustrates the Mn$_{12}$ cluster, where four
inner Mn$^{4+}$ spins and eight outer Mn$^{3+}$ spins are directed
antiparallel to each other and exhibit a novel ground state of total spin
$S=10$ \cite{C5873}.
Assuming the stability of the collective spin of $S=10$, the Mn$_{12}$
cluster is often treated as a single spin-$10$ object \cite{C938}.
However, recent electron-paramagnetic-resonance measurements \cite{H2453}
suggest a possible breakdown of the spin-$10$ description even at low
temperatures.
We have to consider the intracluster magnetic structure for further
understanding of nanoscale magnets.
Since NMR measurements on the Mn$_{12}$ cluster have recently made
remarkable progress \cite{L3773,A064420,F104401,K224425,G104408}, we
consider a microscopic interpretation \cite{Y157603} of them.

   The total spin states in the Mn$_{12}$ cluster is too large even for
modern computers to directly handle.
Indeed several authors \cite{R014408,R054409} have recently succeeded in
obtaining the low-lying eigenstates of the microscopic Hamiltonian, yet
very little is known about the intracluster magnetic structure.
There are four types of exchange interactions between the Mn ions forming
the Mn$_{12}$ cluster, as is illustrated in Fig. \ref{F:Mn12}, but their
values are still the subject of controversy.
Taking account of the current argument, we consider three possible sets of
parameters.
The Hamiltonian is defined as \cite{S1804}
\begin{eqnarray}
   &&
   {\cal H}
   =-\sum_{n=1}^4
    \Bigl[
     2J_1\mbox{\boldmath$s$}_{n}\cdot\mbox{\boldmath$S$}_{n}
    +2J_2(\mbox{\boldmath$s$}_{n}\cdot
          \mbox{\boldmath$\widetilde{S}$}_{n}
         +\mbox{\boldmath$\widetilde{S}$}_{n}\cdot
          \mbox{\boldmath$s$}_{n+1})
   \nonumber \\
   &&\quad
    +2J_3(\mbox{\boldmath$s$}_{n}\cdot\mbox{\boldmath$s$}_{n+1}
         +\frac{1}{2}
          \mbox{\boldmath$s$}_{n}\cdot\mbox{\boldmath$s$}_{n+2})
   \nonumber \\
   &&\quad
    +2J_4(\mbox{\boldmath$S$}_{n}\cdot
          \mbox{\boldmath$\widetilde{S}$}_{n}
         +\mbox{\boldmath$\widetilde{S}$}_{n}\cdot
          \mbox{\boldmath$S$}_{n+1})
   \nonumber \\
   &&\quad
    +D_2(S_{n}^z)^2
    +D_3(\widetilde{S}_{n}^z)^2
    +g\mu_{\rm B}H(s_{n}^z+S_{n}^z+\widetilde{S}_{n}^z)
    \Bigr]\,,
   \label{E:HMn12}
\end{eqnarray}
\noindent
where $\mbox{\boldmath$s$}_n$, $\mbox{\boldmath$S$}_n$, and
$\widetilde{\mbox{\boldmath$S$}}_n$ are the spin operators for the
Mn(1) (spin $\frac{3}{2}\equiv s$), Mn(2) (spin $2\equiv S$), and
Mn(3) (spin $2\equiv S$) sites in the $n$th unit, respectively.
The parameter sets (a), (b), and (c) listed in Table \ref{T:J} are based
on the calculations in Refs. \cite{Z1140}, \cite{S1804}, and
\cite{R054409}, respectively.
The early arguments (a) and (b) agree on $J_1$ predominating over the
rest, which is supported by recent calculations \cite{Y157603,R014408},
whereas the latest claim (c) is completely against such a consensus,
throwing doubt on previous investigations such as the eight-spin modeling
of the Mn$_{12}$ cluster \cite{K6919}.
As for the anisotropy parameters, there is much less information.
When the molecule is treated as a rigid spin-$10$ object, the macroscopic
uniaxial crystalline anisotropy parameter $D$ is determined so as to fit
the zero-field separation between the lowest two levels,
${\mit\Delta}\simeq 14\,\mbox{K}$ \cite{C5873,S1804}.
Hence it is natural to choose the local single-ion anisotropy parameters
$D_2$ and $D_3$, describing the Jahn-Teller-distorted Mn$^{3+}$ ions
\cite{S141}, within the same scheme.

   We introduce the bosonic operators within the linear spin-wave scheme
as $s_n^z=-s+a_{n,1}^\dagger a_{n,1}$, $s_l^+=\sqrt{2s}a_{n,1}^\dagger$;
$S_n^z=S-a_{n,2}^\dagger a_{n,2}$, $S_n^+=\sqrt{2S}a_{n,2}$;
$\widetilde{S}_n^z=S-a_{n,3}^\dagger a_{n,3}$,
$\widetilde{S}_n^+=\sqrt{2S}a_{n,3}$,
which can be justified in the low-temperature range where $^{55}$Mn NMR
measurements \cite{F104401,K224425,G104408} are performed.
While the same type constraint as Eq. (\ref{E:Mst:}) is given by
\begin{equation}
   \sum_k
   \sum_{n=1,2,3}
    \bar{n}_{k,n}
   \sum_{m=1,2,3}
    |\psi_{mn}(k)|^2
   =8S+4s\,,
   \label{E:Mst:Mn12}
\end{equation}
providing we define the Bogoliubov transformation as
\begin{equation}
   \left.
   \begin{array}{rrrr}
    a_{k,1}=&-\psi_{11}  (k)\,b_{k,1}^\dagger
            &-\psi_{12}  (k)\,b_{k,2}^\dagger
            &+\psi_{13}  (k)\,b_{k,3}\,, \\
    a_{k,2}=& \psi_{21}^*(k)\,b_{k,1}
            &+\psi_{22}^*(k)\,b_{k,2}
            &-\psi_{23}^*(k)\,b_{k,3}^\dagger\,, \\
    a_{k,3}=& \psi_{31}^*(k)\,b_{k,1}
            &+\psi_{32}^*(k)\,b_{k,2}
            &-\psi_{33}^*(k)\,b_{k,3}^\dagger\,, \\
   \end{array}
   \right.
\end{equation}
this is not applicable to the Mn$_{12}$ cluster as it is.
In isotropic ferrimagnets, there exists a zero-energy excitation and
therefore a certain number of bosons naturally survive at low
temperatures.
Once a gap $\mit\Delta$ opens, which is the case of the anisotropic
Mn$_{12}$ clusters, the boson number should exponentially decreases as
$\propto{\rm e}^{-{\mit\Delta}/k_{\rm B}T}$ at low temperatures,
but the constraint (\ref{E:Mst:Mn12}) still keeps it finite even at
$T\rightarrow 0$.
Hence we replace Eq. (\ref{E:Mst:Mn12}) by
\widetext
\begin{figure}
\centerline
{\mbox{\psfig{figure=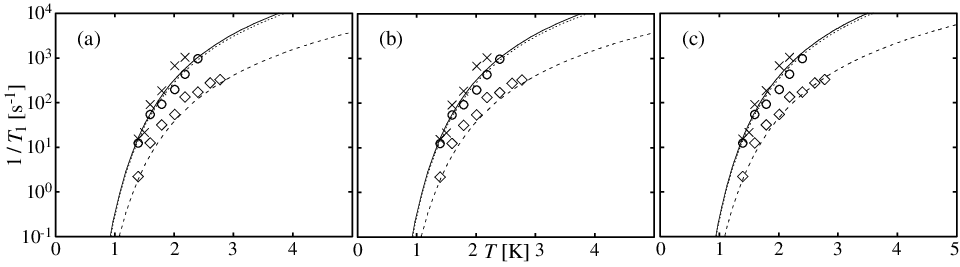,width=160mm,angle=0}}}
\vspace*{2mm}
\caption{Semilog plots of the $^{55}$Mn nuclear spin-lattice relaxation
         rates as functions of temperature under no field for the
         parameter sets listed in Table \ref{T:J}.
         The linear modified spin-wave calculations, dashed [Mn(1)],
         dotted [Mn(2)], and solid [Mn(3)] lines, are compared with
         measurements [67], $\diamond$ [Mn(1)], $\circ$ [Mn(2)], and
         $\times$ [Mn(3)].}
\label{F:Mn12T1T}
\vspace*{8mm}
\centerline
{\mbox{\psfig{figure=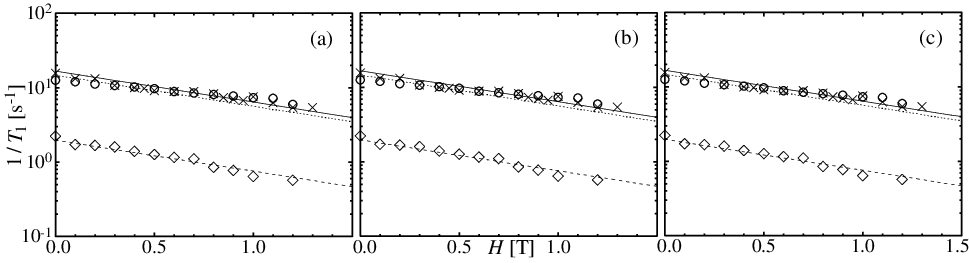,width=160mm,angle=0}}}
\vspace*{2mm}
\caption{Semilog plots of the $^{55}$Mn nuclear spin-lattice relaxation
         rates as functions of an applied field at $T=1.4$ K for the
         parameter sets listed in Table \ref{T:J}.
         The linear modified spin-wave calculations, dashed [Mn(1)],
         dotted [Mn(2)], and solid [Mn(3)] lines, are compared with
         measurements [67], $\diamond$ [Mn(1)], $\circ$ [Mn(2)], and
         $\times$ [Mn(3)].}
\label{F:Mn12T1H}
\end{figure}
\vspace*{2mm}
\narrowtext
\begin{equation}
   \sum_k
   \sum_{n=1,2,3}
    \bar{n}_{k,n}
   \sum_{m=1,2,3}
    |\psi_{mn}(k)|^2
   =(8S+4s)\,{\rm e}^{-{\mit\Delta}/k_{\rm B}T}\,.
   \label{E:Mst:Mn12ex}
\end{equation}
This is quite natural modification of the theory, because the new
constraint (\ref{E:Mst:Mn12ex}) remains the same as the authorized one
(\ref{E:Mst:Mn12}) except for the sufficiently low-temperature region
$k_{\rm B}T\alt{\mit\Delta}$.
It is also convincing that Eq. (\ref{E:Mst:Mn12ex}) smoothly turns into
Eq. (\ref{E:Mst:Mn12}) as ${\mit\Delta}\rightarrow 0$.

   The thus-calculated Raman relaxation rates for the $^{55}$Mn nuclei as
functions of temperature and an applied field are compared with
measurements \cite{F104401} in Figs. \ref{F:Mn12T1T} and \ref{F:Mn12T1H},
where the coupling constants $A_i$ are the only adjustable parameters in
reproducing both temperature and field dependences and have been chosen as
Table \ref{T:Ai}.
The parameter set (b) best agrees to the experimental findings.
At the same time, we are surprised to learn that the latest argument (c)
is also reasonable.
Since the present calculations are not so quantitative as to distinguish
the two scenarios (b) and (c), the intracluster magnetic structure should
be reexamined in all its aspects.
The assumption of the leading Raman process is less trivial for the
Mn$^{4+}$ ions.
A recent experiment \cite{K224425} suggests that the hyperfine field of
the Mn$^{3+}$ ions is in fact anisotropic and makes a predominant
dipolar contribution, whereas that of the Mn$^{4+}$ ions is isotropic and
results in the Fermi contact.
In this context, it is interesting to compare carefully the
theoretical ($A_i^{\rm th}$) and experimental ($A_i^{\rm ex}$)
findings for the coupling constants.
Assuming the set (b),
$A_1^{\rm th}\simeq 2.5A_1^{\rm ex}$,
$A_2^{\rm th}\simeq 1.7A_2^{\rm ex}$, and
$A_3^{\rm th}\simeq 1.2A_3^{\rm ex}$.
Somewhat larger deviation of the theory from the experiment for $A_1$
implies that the nuclear spin-lattice relaxation on the Mn(1) site
may not be Raman active primarily but be strongly influenced by the
surrounding Mn$^{\rm 3+}$ ions.
The present theory is distinct from the phenomenological interpretation
\cite{L3773} assuming phonons to mediate the relaxation.
We hope microscopic investigations of this kind will contribute toward the
total understanding of mesoscopic magnetism.

\section{Spin-Gapped Antiferromagnets}

   We have fully discussed the systems with macroscopically degenerate
magnetic ground states, where the difficulty of diverging number of bosons
does not occur unless temperature is sufficiently high.
From this point of view, one-dimensional antiferromagnets are harder for
spin waves to treat, where the problem lies already in the ground state.
However, here are not a few interesting systems such as Haldane-gap
antiferromagnets \cite{H464} and antiferromagnetic spin ladders
\cite{D5744}.
The antiferromagnetic modified spin-wave theory is not so successful as
that for ferro- and ferrimagnets, but it can still play a helpful role in
our explorations.
We close our discussion by demonstrating bosonic treatments of
integer-spin Heisenberg chains \cite{Y769}.

   We are expected to avoid the quantum divergence of sublattice
magnetizations maintaining the core idea of constant bosons in number.
We introduce the sublattice bosons as Eq. (\ref{E:HP}), where $S=s$, and
diagonalize an effective Hamiltonian
\begin{equation}
   \widetilde{\cal H}
   ={\cal H}+2J\lambda\sum_n(a_n^\dagger a_n+b_n^\dagger b_n)\,,
   \label{E:effH}
\end{equation}
instead of ${\cal H}$, where the Lagrange multiplier $\lambda$ is
determined by the condition
\begin{equation}
   \sum_n a_n^\dagger a_n=\sum_n b_n^\dagger b_n=SN\,.
   \label{E:MstMSW}
\end{equation}
Within the conventional spin-wave theory, spins on one sublattice point
predominantly up, while those on the other predominantly down.
{\it The constraint (\ref{E:MstMSW}) restores the sublattice symmetry}.

   It is interesting to compare the modified spin-wave scheme with another
bosonic approach.
The Schwinger-boson representation of the spin algebra \cite{A316,S5028}
is also a common language to interpret quantum magnetism and its
application to one-dimensional systems \cite{W1057,C915} is moving ahead
recently.
Let us describe each spin variable in terms of two kinds of bosons as
\begin{equation}
   S_l^+=a_{l\uparrow}^\dagger a_{l\downarrow}\,,\ 
   S_l^-=a_{l\downarrow}^\dagger a_{l\uparrow}\,,\ 
   S_l^z=\frac{1}{2}
         \bigl(a_{l\uparrow}^\dagger a_{l\uparrow}
              -a_{l\downarrow}^\dagger a_{l\downarrow}\bigr)\,,
   \label{E:SB}
\end{equation}
where we impose the constraint
\begin{equation}
   a_{l\uparrow}^\dagger a_{l\uparrow}
  +a_{l\downarrow}^\dagger a_{l\downarrow}=2S\,,
   \label{E:MstSB}
\end{equation}
on the bosons.
At the mean-field level, we diagonalize the Hamiltonian with the
constraints imposed on the average and obtain the dispersion relation as
\begin{equation}
   \omega_k=2\sqrt{\lambda^2-\Omega^2\cos^2(ak)}\,,
   \label{E:dspSB}
\end{equation}
where $\lambda$ and $\Omega$ are determined through
\begin{equation}
   \left.
   \begin{array}{lll}
    {\displaystyle\sum_k
     \frac{2\lambda(2\bar{n}_k+1)}
          {\sqrt{\lambda^2-\Omega^2\cos^2(ak)}}}
     &=&
     (2S+1)L\,,\\
    {\displaystyle\sum_k
     \frac{(2\bar{n}_k+1)\cos^2(ak)}
          {\sqrt{\lambda^2-\Omega^2\cos^2(ak)}}}
     &=&
     L\,,
   \end{array}
   \right.
\end{equation}
with $\bar{n}_k=[{\rm e}^{\omega_k/k_{\rm B}T}-1]^{-1}$.
The magnetic susceptibility is expressed as
\begin{figure}
\centerline
{\mbox{\psfig{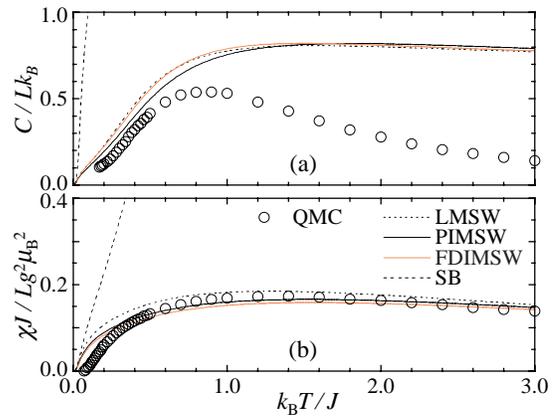}}}
\vspace*{2mm}
\caption{Temperature dependences of the specific heat (a) and the magnetic
         susceptibility (b) for the $S=1$ antiferromagnetic Heisenberg
         chain calculated by the linear modified spin waves (LMSW), the
         perturbational (PIMSW) and full-diagonalization (FDIMSW)
         interacting modified spin waves, the Schwinger bosons (SB), and
         a quantum Monte Carlo method (QMC).}
\label{F:SGFT}
\end{figure}
\vspace*{2mm}
\begin{equation}
   \chi
    =\frac{(g\mu_{\rm B})^2}{k_{\rm B}T}
     \sum_k\sum_{\sigma=\pm}
     \bar{n}_k(\bar{n}_k+1)\,,
   \label{E:chiTSB}
\end{equation}
while the internal energy as
\begin{equation}
   E=JL\bigl[S^2+2\Omega^2-2\lambda(2S+1)\bigr]
  +2\sum_k(2\bar{n}_k+1)\omega_k\,,
   \label{E:ESB}
\end{equation}
where we have corrected the mean-field artifact \cite{A316} of double
counting the degrees of freedom for the Schwinger bosons.

   In Table \ref{T:Haldane} all the present findings for the ground-state
energy and the lowest excitation gap are compared with quantum Monte Carlo
calculations \cite{T047203}.
As for the modified spin-wave calculations, we can treat the $O(S^0)$
correlations in two different ways.
One idea is the perturbational treatment of ${\cal H}_0$ to ${\cal H}_1$,
which is referred to as the perturbational interacting modified spin-wave
scheme, while the other is the full diagonalization of
${\cal H}_1+{\cal H}_0$, which is referred to as the full-diagonalization
interacting modified spin-wave scheme.
On the whole, the bosonic languages well represent the ground-state energy
but fail to describe the Haldane gap quantitatively.
However, they are still useful in understanding the nature of magnetic
excitations.
Although the modified spin waves underestimate the absolute value of
the gap, they reproduce the observed temperature dependence of the Haldane
gap \cite{S3025} better than the nonlinear $\sigma$ model
\cite{Y769,J9265}.
Thus the spin-wave picture still looks valid for spin-gapped
antiferromagnets.
The Haldane-gap phase may be symbolically interpreted as a valence bond
solid \cite{A799}, but such a picture is not strictly valid for the pure
Heisenberg Hamiltonian.
In the case of spin $1$, for example, even a likely ground state of bound
crackions \cite{K627,F8983} moving in the valence-bond-solid background
gives a variational energy \cite{Y157} inferior to the interacting
modified spin-wave estimates.
It is also interesting that the full-diagonalization interacting modified
spin waves and the mean-field Schwinger bosons give the same estimate of
the Haldane gap.
The Schwinger-boson dispersion relation (\ref{E:dspSB}) indeed coincides
exactly with that of the full-diagonalization interacting modified spin
waves at $T=0$.
This is not so surprising because the Holstein-Primakoff bosons are
obtained by replacing both $a_{l\uparrow}$ and $a_{l\uparrow}^\dagger$ by
$(2S-a_{l\downarrow}^\dagger a_{l\downarrow})^{1/2}$ in the
transformation (\ref{E:SB}).

   On the other hand, the modified spin waves are definitely distinguished
from the Schwinger bosons in thermal calculation.
Their findings for the specific heat and the magnetic susceptibility are
compared in Fig. \ref{F:SGFT}.
While the Schwinger-boson mean-field theory rapidly breaks down with
increasing temperature, the modified spin waves still give useful
information at finite temperatures.
The modified spin-wave calculations well reproduce the overall behavior of
the susceptibility converging into
$\chi/Lg^2\mu_{\rm B}^2=S(S+1)/3k_{\rm B}T$ at high temperatures.
We admit, however, that one of the fatal weak points in the modified
spin-wave description of spin-gapped antiferromagnets is nonvanishing
specific heat at high temperatures.
It is not the case with ferro- and ferrimagnets provided we introduce the
Lagrange multiplier in constructing thermodynamics.
The endlessly increasing energy with increasing temperature is because of
the temperature-dependent energy spectrum, where the Lagrange multiplier,
playing the chemical potential, turns out a monotonically increasing
function of temperature.
Such a difficulty is common to all the bosonic approaches to spin-gapped
antiferromagnets, where a fermionic language may be more effective
\cite{D964,H1607,H}.

\section{Concluding Remarks}

   In the late 1980's, the low-dimensional spin-wave theory made an
epochal progress, where a problem pending in one-dimensional ferromagnets
was solved and two-dimensional antiferromagnets were discussed in the
context of high-temperature superconductivity.
More than a decade has passed and the spin-wave theory has grown more
useful and extensive covering one and possibly zero dimensions.
There are further interesting investigations in terms of spin waves for
ladder \cite{H,I144429,N1380}, frustrated \cite{I14456}, and random-bond
\cite{W014429} systems.
The present findings can be reconstructed through the Dyson-Maleev
transformation \cite{D1217,M776} instead of the Holstein-Primakoff scheme.
The modified spin-wave theory breaks the rotational symmetry anyway, while
the Schwinger-boson representation \cite{A316,S5028,W1057} is rotationally
invariant.

   Considering the rich harvest \cite{T168,Y14008,Y157603} in the ferro-
and ferrimagnetic modified spin-wave theory, the antiferromagnetic modified
spin-wave scheme \cite{T2494,H4769,Y769} is less successful in one
dimension in particular, where the ground-state properties and the
magnetic susceptibility are well calculated, while the energy gap and the
specific heat are poorly reproduced.
Recently the Jordan-Wigner spinless fermions defined along a snake-like
path \cite{D964,H1607,H} gave a fine description of ladder
antiferromagnets, a piece of which is listed in Table \ref{T:ladder}.
We learn that the spin-gap is better reproduced by the spinless fermions.
The bosonic and fermionic languages are complementary in our explorations
of low-dimensional magnets.
All these arguments stimulate further interest \cite{K104427} in the
spin-wave theory.
The spin-wave scheme will be refined more and more in the 21st century.

\acknowledgments

It would be hard to list all colleagues with whom I had fruitful
discussion and collaboration.
I am in particular grateful to Professor T. Goto, Professor T. Fukui, and
Dr. N. Fujiwara for invaluable discussions.
I acknowledge financial supports by the Ministry of Education, Culture,
Sports, Science, and Technology of Japan,
the Sumitomo Foundation, and the Nissan Science Foundation.

\begin{table}
\caption{Possible exchange interactions for the Mn$_{12}$ cluster in the
         unit of $\mbox{cm}^{-1}$.}
\begin{tabular}{crrrr}
 & $J_1$ & $J_2$ & $J_3$ & $J_4$ \\
\tableline
\noalign{\vskip 1mm}
(a) & $-150$ & $-60$ & $ 60$ & $ 30$ \\
(b) & $-150$ & $-60$ & $-30$ & $ 30$ \\
(c) & $- 41$ & $-41$ & $  3$ & $- 8$ \\
\end{tabular}
\label{T:J}
\end{table}

\begin{table}
\caption{Estimates of the hyperfine interaction
         $g\mu_{\rm B}\gamma_{\rm N}A_i$ in the unit of
         $\mbox{rad}\cdot\mbox{Hz}$
         between the nuclear and electronic spins on the Mn($i$) site.
         The linear modified spin-wave calculations are compared with
         experimental findings [67].}
\begin{tabular}{lccc}
 & Mn(1) & Mn(2) & Mn(3) \\
\noalign{\vskip 1mm}
\tableline
\noalign{\vskip 1mm}
Theory (a) & $8.9\times 10^{8}$
           & $1.8\times 10^{9}$
           & $1.9\times 10^{9}$ \\
Theory (b) & $2.5\times 10^{8}$
           & $5.1\times 10^{8}$
           & $5.4\times 10^{8}$ \\
Theory (c) & $3.3\times 10^{8}$
           & $5.1\times 10^{8}$
           & $6.6\times 10^{8}$ \\
Experiment & $1.0\times 10^{8}$
           & $3.0\times 10^{8}$
           & $4.5\times 10^{8}$ \\
\end{tabular}
\label{T:Ai}
\end{table}

\begin{table}
\caption{The ground-state energy per spin $E_{\rm g}/L$ and the
         lowest excitation gap ${\mit\Delta}$ for the $S=1$
         antiferromagnetic Heisenberg chain calculated by the linear
         modified spin waves (LMSW), the perturbational (PIMSW) and
         full-diagonalization (FDIMSW) interacting modified spin waves,
         the Schwinger bosons at the mean-field level (SB), and a quantum
         Monte Carlo method (QMC) [84].}
\begin{tabular}{lll}
 & \quad\ $E_{\rm g}/L$ & \quad\ ${\mit\Delta}$ \\
\tableline
\noalign{\vskip 1mm}
LMSW   &$-1.361879   $&$0.07200   $\\
PIMSW  &$-1.394853   $&$0.07853   $\\
FDIMSW &$-1.394617   $&$0.08507   $\\
SB     &$-1.396148   $&$0.08507   $\\
QMC    &$-1.401481(4)$&$0.41048(6)$\\
\end{tabular}
\label{T:Haldane}
\end{table}

\begin{table}
\caption{The ground-state energy per rung $E_{\rm g}/N$ and the
         lowest excitation gap ${\mit\Delta}$ for the $S=\frac{1}{2}$
         antiferromagnetic Heisenberg two-leg ladder with intra- and
         interchain couplings of equal strength calculated by the
         spinless fermions within the Hartree (HSF) and Hartree-Fock
         (HFSF) approximations, the linear modified spin waves (LMSW),
         the perturbational interacting modified spin waves (PIMSW), and
         a density-matrix renormalization-group method (DMRG) [99].}
\begin{tabular}{lll}
 & \quad\ $E_{\rm g}/L$ & \quad\ ${\mit\Delta}$ \\
\noalign{\vskip 1mm}
\tableline
\noalign{\vskip 1mm}
HSF  & $-0.838805$ & $0.500000$ \\
HFSF & $-0.907183$ & $0.450419$ \\
LMSW & $-1.101403$ & $0.100911$ \\
IMSW & $-1.143750$ & $0.112832$ \\
DMRG & $-1.156086$ & $0.504$    \\
\end{tabular}
\label{T:ladder}
\end{table}

\widetext

\begin{references}
\bibitem{B206}
   F. Bloch,
      Z. Phys. {\bf 61}, 206 (1930).

\bibitem{H1098}
   T. Holstein and H. Primakoff,
      Phys. Rev. {\bf 58}, 1098 (1940).

\bibitem{D1217}
   F. J. Dyson,
      Phys. Rev. {\bf 102}, 1217 (1956);
                            1230 (1956).

\bibitem{O117}
   T. Oguchi,
      Phys. Rev. {\bf 117}, 117 (1960).

\bibitem{K25}
   F. Keffer and R. Loudon,
      J. Appl. Phys. {\bf 32}, 25 (1961);
                     {\bf 33}, 250 (1962).

\bibitem{K1384}
   J. Kanamori and M. Tachiki,
      J. Phys. Soc. Jpn. {\bf 17}, 1384 (1962).

\bibitem{A694}
   P. W. Anderson,
      Phys. Rev. {\bf 86}, 694 (1952).

\bibitem{K568}
   R. Kubo,
      Phys. Rev. {\bf 87}, 568 (1952).

\bibitem{G122}
   A. C. Gossard, V. Jaccarino, and J. P. Remeika,
      Phys. Rev. Lett. {\bf 7}, 122 (1961).

\bibitem{W1357}
   R. Weber and P. E. Tannenwald,
      J. Phys. Chem. Solids {\bf 24}, 1357 (1963).

\bibitem{N1929}
   A. Narath,
      Phys. Rev. {\bf 131}, 1929 (1963).

\bibitem{D433}
   H. L. Davis and A. Narath,
      Phys. Rev. {\bf 134}, A433 (1964);
                 {\bf 137}, A163 (1965).

\bibitem{M23}
   T. Moriya,
      Prog. Theor. Phys. {\bf 16}, 23 (1956);
                                  641 (1956).
\bibitem{P398}
   P. Pincus,
      Phys. Rev. Lett. {\bf 16}, 398 (1966);
   D. Beeman and P. Pincus,
      Phys. Rev. {\bf 166}, 359 (1968).

\bibitem{N354}
   A. Narath and A. T. Fromhold, Jr.,
      Phys. Rev. Lett. {\bf 17}, 354 (1966).

\bibitem{K357}
   N. Kaplan, R. Loudon, V. Jaccarino, H. J. Guggenheim,
   D. Beeman, and P. A. Pincus:
      Phys. Rev. Lett. {\bf 17}, 357 (1966).

\bibitem{W1055}
   R. D. Willett, C. P. Landee, R. M. Gaura, D. D. Swank,
   H. A. Groenendijk, and A. J. van Duyneveldt,
      J. Magn. Magn. Mater. {\bf 15}-{\bf 18}, 1055 (1980).

\bibitem{T2808}
   M. Takahashi and M. Yamada,
      J. Phys. Soc. Jpn. {\bf 54}, 2808 (1985);
   M. Yamada and M. Takahashi,
      J. Phys. Soc. Jpn. {\bf 55}, 2024 (1986).

\bibitem{S2131}
   P. Schlottmann,
      Phys. Rev. Lett. {\bf 54}, 2131 (1985).

\bibitem{B640}
   J. C. Bonner and M. E. Fisher,
      Phys. Rev. {\bf 135}, A640 (1964).

\bibitem{B1272}
   G. A. Baker, Jr., G. S. Rushbrooke, and H. E. Gilbert,
      Phys. Rev. {\bf 135}, A1272 (1964).

\bibitem{K807}
   J. Kondo and K. Yamaji,
      Prog. Theor. Phys. {\bf 47}, 804 (1972).

\bibitem{C297}
   J. J. Cullen and D. P. Landau,
      Phys. Rev. B {\bf 27}, 297 (1983).

\bibitem{L3108}
   J. W. Lyklema,
      Phys. Rev. B {\bf 27}, 3108 (1983).

\bibitem{T233}
   M. Takahashi,
      Prog. Theor. Phys. Suppl. {\bf 87}, 233 (1986).

\bibitem{T168}
   M. Takahashi,
      Phys. Rev. Lett. {\bf 58}, 168 (1987).

\bibitem{T2494}
   M. Takahashi,
      Phys. Rev. B {\bf 40}, 2494 (1989).

\bibitem{H4769}
   J. E. Hirsch and S. Tang,
      Phys. Rev. B {\bf 40}, 4769 (1989);
   S. Tang, M. E. Lazzouni, and J. E. Hirsch,
      Phys. Rev. B {\bf 40}, 5000i1989).

\bibitem{C7832}
   H. A. Ceccatto, C. J. Gazza, and A. E. Trumper,
      Phys. Rev. B {\bf 45}, 7832 (1992).

\bibitem{D13821}
   A. V. Dotsenko and O. P. Sushkov,
      Phys. Rev. B {\bf 50}, 13821 (1994).

\bibitem{R2589}
   S. M. Rezende,
      Phys. Rev. B {\bf 42}, 2589 (1990).

\bibitem{M1}
   E. Manousakis,
      Rev. Mod. Phys. {\bf 63}, 1 (1991).

\bibitem{Y1024}
   S. Yamamoto,
      Phys. Rev. B {\bf 59}, 1024 (1999).

\bibitem{B3921}
   S. Brehmer, H.-J. Mikeska, and S. Yamamoto,
      J. Phys.: Condens. Matter {\bf 9}, 3921 (1997).

\bibitem{K3336}
   A. K. Kolezhuk, H.-J. Mikeska, and S. Yamamoto,
      Phys. Rev. B {\bf 55}, R3336 (1997).

\bibitem{Y211}
   S. Yamamoto, T. Fukui, and T. Sakai,
      Eur. Phys. J. B {\bf 15}, 211 (2000).

\bibitem{W433}
   Y. L. Wang and H. B. Callen,
      Phys. Rev. {\bf 148}, 433 (1966).

\bibitem{Y14008}
   S. Yamamoto and T. Fukui,
      Phys. Rev. B {\bf 57}, R14008 (1998);
   S. Yamamoto, T. Fukui, K. Maisinger, and U. Schollw\"ock,
      J. Phys.: Condens. Matter {\bf 10}, 11033 (1998).

\bibitem{K782}
   O. Kahn, Y. Pei, M. Verdaguer, J.-P. Renard, and J. Sletten,
      J. Am. Chem. Soc. {\bf 110}, 782 (1988);
   P. J. van Koningsbruggen, O. Kahn, K. Nakatani, Y. Pei,
   J.-P. Renard, M. Drillon, and P. Legoll,
      Inorg. Chem. {\bf 29}, 3325 (1990).

\bibitem{O5221}
   A. S. Ovchinnikov, I. G. Bostrem, V. E. Sinitsyn, N. V. Baranov, and
   K. Inoue,
      J. Phys.: Condens. Matter {\bf 13}, 5221 (2001).

\bibitem{O8067}
   A. S. Ovchinnikov, I. G. Bostrem, V. E. Sinitsyn,
   A. S. Boyarchenkov, N. V. Baranov, and K. Inoue,
      J. Phys.: Condens. Matter {\bf 14}, 8067 (2002).

\bibitem{M5908}
   K. Maisinger, U. Schollw\"ock, S. Brehmer, H.-J. Mikeska, and
   S. Yamamoto,
      Phys. Rev. B {\bf 58}, R5908 (1998).

\bibitem{D10992}
   M. Drillon, E. Coronado, R. Georges, J. C. Gianduzzo, and J. Curely,
      Phys. Rev. B {\bf 40}, 10992 (1989).

\bibitem{C56}
   A. Caneschi, D. Gatteschi, P. Rey, and R. Sessoli,
      Inorg. Chem. {\bf 27}, 1756 (1988);
   A. Caneschi, D. Gatteschi, J.-P. Renard, P. Rey, and R. Sessoli,
      {\it ibid.} {\bf 28}, 1976 (1989); 2940 (1989).

\bibitem{D83}
   M. Drillon, M. Belaiche, P. Legoll, J. Aride, A. Boukhari, and
   A. Moqine,
      J. Magn. Magn. Mater. {\bf 128}, 83 (1993).

\bibitem{E4466}
   A. Escuer, R. Vicente, M. S. El Fallah, M. A. S. Goher, and
   F. A. Mautner,
      Inorg. Chem. {\bf 37}, 4466 (1998).

\bibitem{N214418}
   T. Nakanishi and S. Yamamoto,
      Phys. Rev. B {\bf 65}, 214418 (2002).

\bibitem{H30}
   M. Hagiwara, Y. Narumi, K. Minami, and K. Kindo,
      Physica B {\bf 294}-{\bf 295}, 30 (2001).

\bibitem{Y13610}
   S. Yamamoto, S. Brehmer, and H.-J. Mikeska,
      Phys. Rev. B {\bf 57}, 13610 (1998).

\bibitem{Y842}
   S. Yamamoto,
      Phys. Rev. B {\bf 61}, R842 (2000);
      Phys. Lett. A {\bf 265}, 139 (2000).

\bibitem{F433}
   N. Fujiwara and M. Hagiwara,
      Solid State Commun. {\bf 113}, 433 (2000).

\bibitem{Y2324}
   S. Yamamoto,
      J. Phys. Soc. Jpn. {\bf 69}, 2324 (2000).

\bibitem{Y1}
   S. Yamamoto,
      Solid State Commun. {\bf 117}, 1 (2001).

\bibitem{H965}
   D. Hone, C. Scherer, and F. Borsa,
      Phys. Rev. B {\bf 9}, 965 (1974);
   F. Borsa and M. Mali,
      Phys. Rev. B {\bf 9}, 2215 (1974);
   J.-P. Boucher, M. A. Bakheit, M. Nechtschein, M. Villa, G. Bonera, and
   F. Borsa,
      Phys. Rev. B {\bf 13}, 4098 (1976).

\bibitem{T2173}
   M. Takigawa, T. Asano, Y. Ajiro, M. Mekata, and Y. J. Uemura,
      Phys. Rev. Lett. {\bf 76}, 2173 (1996).

\bibitem{F11945}
   N. Fujiwara, H. Yasuoka, M. Isobe, Y. Ueda, and S. Maegawa,
      Phys. Rev. B {\bf 55}, R11945 (1997).

\bibitem{C661}
   E. M. Chudnovsky and L. Gunther,
      Phys. Rev. Lett. {\bf 60}, 661 (1988).

\bibitem{H0544XX}
   H. Hori and S. Yamamoto,
      Phys. Rev. B {\bf 68}, 0544XX (2003).

\bibitem{L2042}
   T. Lis,
      Acta Crystallogr. Sect. B {\bf 36}, 2042 (1980).

\bibitem{F3830}
   J. R. Friedman, M. P. Sarachik, J. Tejada, and R. Ziolo,
      Phys. Rev. Lett. {\bf 76}, 3830 (1996).

\bibitem{T145}
   L. Thomas, F. Lionti, R. Ballou, D. Gatteschi, R. Sessoli, and
   B. Barbara,
      Nature {\bf 383}, 145 (1996).

\bibitem{C5873}
   A. Caneschi, D. Gatteschi, R. R. Sessoli, A. L. Barra,
   L. C. Brunel, and M. Guillot,
      J. Am. Chem. Soc. {\bf 113}, 5873 (1991).

\bibitem{C938}
   E. M. Chudnovsky,
      Science {\bf 274}, 938 (1996).

\bibitem{H2453}
   S. Hill, J. A. A. J. Perenboom, N. S. Dalal, T. Hathaway,
   T. Stalcup, and J. S. Brooks,
      Phys. Rev. Lett. {\bf 80}, 2453 (1998).

\bibitem{L3773}
   A. Lascialfari, Z. H. Jang, F. Borsam, P. Carretta, and D. Gatteschi,
   Phys. Rev. Lett. {\bf 81}, 3773 (1998).

\bibitem{A064420}
   R. M. Achey, P. L. Kuhns, A. P. Reyes, W. G. Moulton, and N. S. Dalal,
      Phys. Rev. B {\bf 64}, 064420 (2001).

\bibitem{F104401}
   Y. Furukawa, K. Watanabe, K. Kumagai, F. Borsa, and D. Gatteschi,
      Phys. Rev. B {\bf 64}, 104401 (2001).

\bibitem{K224425}
   T. Kubo, T. Goto, T. Koshiba, K. Takeda, and K. Awaga,
      Phys. Rev. B {\bf 65}, 224425 (2002).

\bibitem{G104408}
   T. Goto, T. Koshiba, T. Kubo, and K. Awaga,
      Phys. Rev. B {\bf 67}, 104408 (2003).

\bibitem{Y157603}
   S. Yamamoto and T. Nakanishi,
      Phys. Rev. Lett. {\bf 89}, 157603 (2002).

\bibitem{R014408}
   I. Rudra, S. Ramasesha, and D. Sen,
      Phys. Rev. B {\bf 64}, 014408 (2001);
   C. Raghu, I. Rudra, D. Sen, and S. Ramasesha,
      {\it ibid.} {\bf 64}, 064419 (2001).

\bibitem{R054409}
   N. Regnault, Th. Jolic{\oe}ur, R. Sessoli, D. Gatteschi, and
   M. Verdaguer,
      Phys. Rev. B {\bf 66}, 054409 (2002).

\bibitem{S1804}
   R. Sessoli, H.-L. Tsai, A. R. Schake, S. Wang, J. B. Vincent,
   K. Folting, D. Gatteschi, G. Christou, and D. N. Hendrickson,
      J. Am. Chem. Soc. {\bf 115}, 1804 (1993).

\bibitem{Z1140}
   A. K. Zvezdin and A. I. Popov,
      JETP {\bf 82}, 1140 (1996).

\bibitem{K6919}
   M. I. Katsnelson, V. V. Dobrovitski, and B. N. Harmon,
      Phys. Rev. B {\bf 59}, 6919 (1999).

\bibitem{S141}
   R. Sessoli, D. Gatteschi, A. Caneschi, and M. A. Novak,
      Nature {\bf 365}, 141 (1993).

\bibitem{H464}
   F. D. M. Haldane,
      Phys. Lett. {\bf 93A}, 464 (1983);
      Phys. Rev. Lett. {\bf 50}, 1153 (1983).

\bibitem{D5744}
   E. Dagotto, J. Riera, and D. Scalapino,
      Phys. Rev. B {\bf 45}, R5744 (1992).

\bibitem{Y769}
   S. Yamamoto and H. Hori,
      J. Phys. Soc. Jpn. {\bf 72}, 769 (2003).

\bibitem{A316}
   D. P. Arovas and A. Auerbach,
      Phys. Rev. B {\bf 38}, 316 (1988);
   A. Auerbach and D. P. Arovas,
      Phys. Rev. Lett. {\bf 61}, 617 (1988).

\bibitem{S5028}
   S. Sarker, C. Jayaprakash, H. R. Krishnamurthy, and M. Ma,
      Phys. Rev. B {\bf 40}, 5028 (1989).

\bibitem{W1057}
   C. Wu, B. Chen, X. Dai, Y. Yu, and Z.-B. Su,
      Phys. Rev. B {\bf 60}, 1057 (1999).

\bibitem{C915}
   X. Y. Chen, Q. Jiang, and W. Z. Shen,
      J. Phys.: Condens. Matter {\bf 15} (2003) 915.

\bibitem{T047203}
   S. Todo and K. Kato,
      Phys. Rev. Lett. {\bf 87}, 047203 (2001).

\bibitem{S3025}
   T. Sakaguchi, K. Kakurai, T. Yokoo, and J. Akimitsu,
      J. Phys. Soc. Jpn. {\bf 65}, 3025 (1996).

\bibitem{J9265}
   Th. Jolic{\oe}ur and O. Golinelli,
      Phys. Rev. B {\bf 50}, 9265 (1994).

\bibitem{A799}
   I. Affleck, T. Kennedy, E. H. Lieb, and H. Tasaki,
      Phys. Rev. Lett. {\bf 59}, 799 (1987);
      Commun. Math. Phys. {\bf 115}, 477 (1988).

\bibitem{K627}
   S. Knabe,
      J. Stat. Phys. {\bf 52}, 627 (1988).

\bibitem{F8983}
   G. F\'ath and J. S\'olyom,
      J. Phys.: Condens. Matter {\bf 5}, 8983 (1993).

\bibitem{Y157}
   S. Yamamoto,
      Phys. Lett. A. {\bf 225}, 157 (1997);
      Int. J. Mod. Phys. B {\bf 12}, 1795 (1998).

\bibitem{D964}
   X. Dai and Z.-B. Su,
      Phys. Rev. B {\bf 57}, 964 (1998).

\bibitem{H1607}
   H. Hori and S. Yamamoto,
      J. Phys. Soc. Jpn. {\bf 71}, 1607 (2002).

\bibitem{H}
   H. Hori and S. Yamamoto,
      preprint.

\bibitem{I144429}
   N. B. Ivanov and J. Richter,
      Phys. Rev. B {\bf 63}, 144429 (2001).

\bibitem{N1380}
   T. Nakanishi, S. Yamamoto, and T. Sakai,
      J. Phys. Soc. Jpn. {\bf 70}, 1380 (2001).

\bibitem{I14456}
   N. B. Ivanov, J. Richter, and U. Schollw\"ock,
      Phys. Rev. B {\bf 58}, 14456 (1998).

\bibitem{W014429}
   X. Wan, K. Yang, and R. N. Bhatt,
      Phys. Rev. B {\bf 66}, 014429 (2002).

\bibitem{M776}
   S. V. Maleev,
      Sov. Phys. JETP {\bf 6}, 776 (1958).

\bibitem{W886}
   S. R. White, R. M. Noack, D. J. Scalapino,
      Phys. Rev. Lett. {\bf 73} (1994) 886.

\bibitem{K104427}
   M. Kollar, I. Spremo, and P. Kopietz,
      Phys. Rev. B {\bf 67}, 104427 (2003).

\end{references}
\end{document}